# Development of a Chinese Human-Automation Trust Scale

Zixin Cui[1]*, Xiangling Zhuang[2], Seul Chan Lee[3], Jieun Lee[4], Xintong Li[1], Makoto Itoh[5, 6]

[1] Department of Risk Engineering, Graduate School of System and Information Engineering, University of Tsukuba, Tsukuba, Ibaraki, Japan

[2] Shaanxi Key Laboratory of Behavior and Cognitive Neuroscience, School of Psychology, Shaanxi Normal University, China

[3] Department of Industrial and Systems Engineering, Gyeongsang National University, Jinju, Republic of Korea

[4] Department of Safety Engineering, Pukyong National University, Busan, Republic of Korea

[5] Institute of Systems and Information Engineering, University of Tsukuba, Tsukuba, Ibaraki, Japan

[6] Center for Artificial Intelligence Research, University of Tsukuba, Tsukuba, Ibaraki, Japan.

*cui@css.risk.tsukuba.ac.jp(ZC)

# Abstract

The development of a reliable and valid assessment tool of human-automation trust is an important topic. This study aimed to develop a Chinese version of human-automation trust scale (C-HATS) with reasonable reliability and validity based on Lee and See (2004)'s trust model. After three phases of assessments including exploratory factor analysis, item analysis, and confirmatory factor analysis, different dimensions and items were considered for initial and post-task human-automation trust. For post-task trust, the scale had three dimensions and 11 items and reflected Lee and See (2004)'s model, whereas different from Lee and See (2004)'s model, the final scale had 14 items but only two dimensions for initial trust. Nevertheless, for both initial and post-task trust, reasonable reliability and validity of the scale were verified with various consumer automation products. Although further verification is still necessary, the developed C-HATS could be used to effectively assess human-automation trust in the Chinese context.

# 1. Introduction

## *1.1. Background*

Human-automation trust has been widely discussed in several areas, such as driving (Payre et al., 2016; Forster et al., 2018), aircraft (e.g., Endsley, 2018), and medicine (e.g., Holzinger and Muller, 2021). According to Lee and See (2004), human-automation trust refers to *the attitude that an automation agent will help achieve an individual's goals in a situation characterized by uncertainty and vulnerability*. Human-automation trust has an important relationship with not only the social acceptance of automation (Hutson, 2017; Adnan et al., 2018), but also the safety improvement of human-automation interaction (Kaber, & Endsley, 1997; Payre et al., 2016). For social acceptance, Zhang et al. (2021) proposed an extended automated vehicles acceptance model and found that trust contributed most among several determinants of the public's acceptance of automated vehicles. The higher the level of trust, the higher the level of social acceptance. For safety improvement, overtrust in automation can also lead to misuse (Parasuraman & Riley, 1997) and is thus positively related to operator failures and intervention failures when it is necessary (Kaber & Endsley, 1997). For example, drivers' overtrust in driving automation can lead to a longer reaction time when drivers were required to cope with an emergence manual control from automated driving systems (Payre et al., 2016). Accordingly, the level of trust can affect human-automation interaction.

Maintaining a suitable level of human-automation trust relies on a reliable and valid measurement of human-automation trust. The widely used measurement is Jian et. al. (2000)'s empirically driven scale with trust and distrust dimensions, which was developed by a word elicitation study and the factor analysis and cluster analyses of the elicited words related to trust or distrust. Subsequent studies (e.g., Bisantz & Seong, 2001; Spain et al., 2008) have verified the validity and reliability of this scale. However, as this scale did not directly reflect any human-automation trust theories, practical implications for human-automation interaction based on empirical findings using this scale may not necessarily rest on the theoretical foundations and lack of robustness (Yamani et al., under review). A reliable and valid theoretically driven scale is thus necessary. The following sections will make a literature review of the existing theoretically driven scale of human-automation trust, and this study attempts to fill the gap of the Lack of a reliable and valid theoretically driven trust scale in the Chinese background.

## *1.2. Literature review of theoretically driven scale*
### *1.2.1. Theoretical model of human-automation trust*

Several theoretical models have been established to discuss the structure of human-automation trust. For example, Muir and Moray (1996) initially proposed a cross-sectional model indicating three dimensions of human-automation trust, *predictability*, *dependability*, and *faith*. Additionally, Hoff and Bashir (2015) proposed a longitudinal three-layered framework illustrating the evolving nature of human-automation trust over time. Presently, one of the widely acknowledged models is proposed by Lee and See (2004). Combining Muir and Moray (1996)'s model and other studies on the basis of human trust (e.g., Barber, 1983; Mayer et al., 1995; Rempel et al., 1985), Lee and See (2004) updated the three dimensions of human-automation trust to *performance*, *process*, and *purpose*.

*Performance* focuses on the specific behaviors of an automation system in the current or the past and involves the perceived *reliability* (Rempel et al., 1985), *predictability* (Muir & Moray,1996), and *ability* (Deutsch, 1960; Muir & Moray,1996) of the system. *Process* involves the inherent qualities and characteristics of the system with which the algorithms are appropriate for the usage situation and can achieve operators' usage purposes of the system (Lee and See, 2004). Therefore, *process* is close or related to *understandability* (forming a mental model and predicting the behaviors of the automation system (Sheridan, 1988)), *dependability* (the extent can count on the automation system to do its job) in Muir and Moray (1996)'s trust model, and *openness* (mental accessibility and the willingness to share ideas and information freely (Butler & Cantrell, 1984)) in interpersonal trust. *Purpose* is related to whether the automation system is used within the design intents and corresponding to *faith* (the degree to which the operator believes the system will be able to cope with other system states in the future, Muir & Moray,1996) and *benevolence* (the extent to which the trustee is believed to want to do good to the trustor in interpersonal relationships, Mayer et al., 1995) that can reflect trustors' perceptions of a positive orientation of the trustee.

### 1.2.2. Previous trust scales based on Lee and See (2004)'s model

As Lee & See's model has been widely recognized, scales to evaluate human-automation trust have been developed (Chancey et al., 2017; Chien et al., 2014; and Korber, 2019) based on the model **(Figure 1)**. However, each scale has its limitations of assessing human-automation trust, which will be discussed in the following paragraphs.

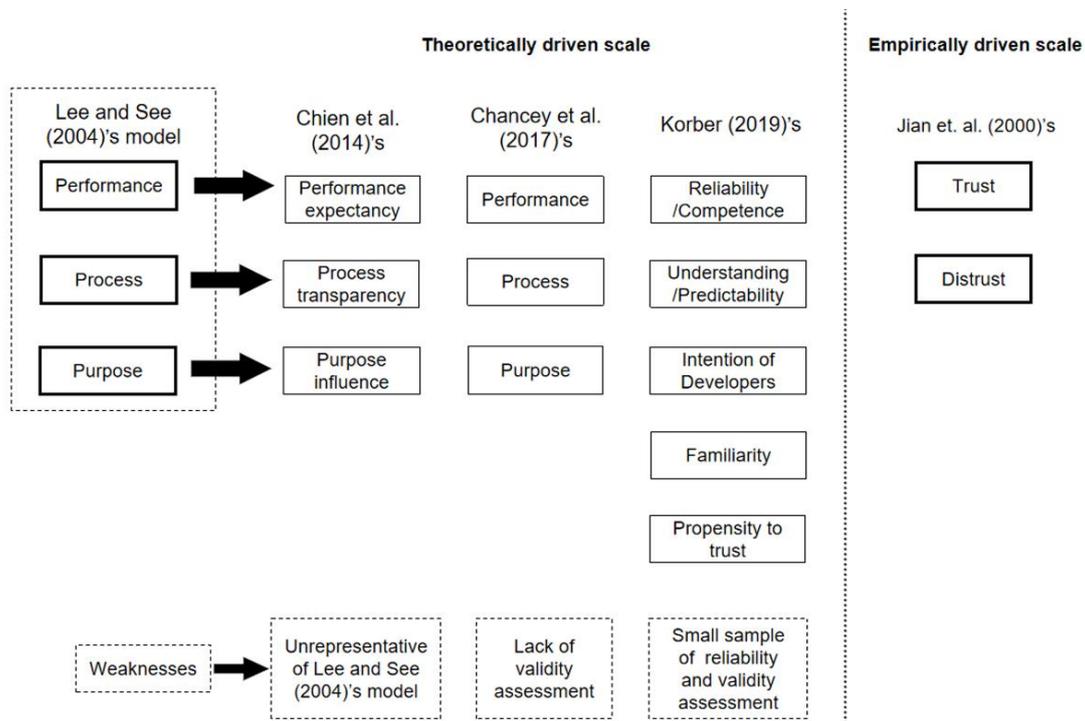

**Figure 1** Theoretical sources of previous human-automation trust scale and Lee and See (2004)'s model.

Chien et al. (2014) have developed a scale for initial trust in automation with good reliability and validity. They at first generated a question pool with 110 items. After exploratory factor analysis, 59 items were retained and three factors of human-automation trust were found, which were named as *performance expectancy*, *process transparency*, and *purpose influence*. According to Chien et al., (2014), *performance expectancy* is a belief that a system will help to enhance operator's job performance, *process transparency* is related to encountered difficulties based on perceived transparency of the system, and *purpose influence* means operator's knowledge of what the system is supposed to do. In addition, some items of the scale seem to be unrepresentative of the factor, such as the item "I am satisfied with the information that a system provides" of *process transparency*. However, the definitions of the three factors are to some extent inconsistent with and are relatively incomplete in comparison to Lee & See (2004)'s. For example, Chien et al., (2014)'s *performance expectancy* is partly related to the *ability* in Lee and See (2004)'s *performance*, but cannot reflect the *predictability* and *reliability* in Lee and See (2004)'s *performance*. Therefore, Chien et al., (2014)'s scale is needed to be improved.

Then, Chancey et al. (2017)'s scale was developed based on Madsen and Gregor (2000)'s human-computer trust scale. With light revisions, the items of *perceived reliability*, *perceived understandability*, and *faith* in Madsen and Gregor (2000)'s scale were respectively taken to align with the three dimensions of *performance*, *process*, and *purpose*. According to Madsen and Gregor (2000), *perceived reliability* refers to perceived repeated, consistent functioning of the system in the usual sense, *perceived understandability* means that the operators can form a mental model and predict the future behaviors of the system, and *faith* is related to system's future ability even in untried situations. It is exactly that the three dimensions in Madsen and Gregor (2000)'s scale have the overlap with Lee and See (2004)'s three, the overlaps of the dimensions however do not mean that the one can be replaced by another one. For example, *perceived reliability* can exactly be included in *performance*, but it cannot totally reflect *performance*. Further, since Chancey et al. (2017) did not test and report the validity of their scale, the scale still needs further verification.

To make a better theoretical consideration and measurement of human-automation trust, Korber (2019) generated representative items after combining the trust model of Mayer et al. (1995) and Lee and See (2004). The final scale has six underlying dimensions - *reliability/competence*, *understandability/predictability*, *intention of developers*, *propensity to trust*, *familiarity*, and *trust in automation*. According to Korber (2019), *reliability/competence*, *understandability/predictability*, and *intention of developers* respectively respond to *performance*, *process*, and *purpose* in Lee and See (2004)'s model. *Familiarity* and *propensity to trust* are two factors that can influence trust in automation as a moderator of *reliability/competence*, *understandability/predictability*, and *intention of developers* on trust in automation. However, due to the small samples (n = 58) for reliability and validity assessments, it is also necessary to further evaluate the scale with larger samples.

### *1.2.3. Lack of a dynamic trust scale for consumer automation in China*

Besides the mentioned limitations of each trust scale, the Lee and See (2004)'s model itself may need further examination for three reasons. One is that although the model has already been examined a lot on automation systems that are used by professional operators (e.g., Korber, 2019; Schaefer et al., 2016), emerging automation systems such as automated driving system has boomed as consumer products to be used by the general public instead of professional operators. Due to the marketability of consumer product and the unprofessional of the users, whether Lee and See (2004)'s model can interpret the public's trust in different consumer automation product should be examined. To address this issue, the current study aims to include several consumer automation systems as the research targets.

Another is that few studies considered the dynamic properties of human-automation trust while verifying Lee and See (2004)'s model. Human-automation trust can develop as the relationship between the trustor and trustee develops (Muir & Moray, 1996; Lee & See, 2004). Previous theories, such as Gao et al (2021)'s dynamic framework of human trust in driving automation, pointed out that there should be four development stages of human-automation trust. The first stage is d*ispositional trust* that reflects the user's propensity to trust (Merritt & Ilgen, 2008; Hoff & Bashir, 2015) and is influenced by users' early trust-related experiences (Hardin, 1996) and some inherent traits such as personality traits and culture (Hoff & Bashir, 2015). The second one is *initial trust* that develops from *dispositional trust* as soon as users have acquired some knowledge of automation, during which users still have none of the direct interactions with the automation. Therefore, *dispositional trust* and indirect experiences related to the system are the two major factors that influence *initial trust* (Gao et al., 2021). Then based on *initial trust* and perceived real-time automation characteristics and usage situations characteristics, *ongoing trust* develops and can dynamically change during the interaction. Finally, *post-task trust* forms after the interaction and may transform into new prior experiences that can influence *initial trust* in other automation systems (Merritt & Ilgen, 2008; Gao et al., 2021). However, it is still unclear whether the structures of human-automation trust at different development phases should be different and the measurements of human-automation trust at different development phases should also be different.

Further, human-automation trust was mainly discussed in western cultural backgrounds, whereas non-western contexts should also be seriously considered (Lee and See, 2004; Gao et al., 2021). Some cross-cultural studies on human-automation interaction found cultural differences can influence human-automation trust (eg., Hu et al., 2019; Abbass et al., 2018). For example, people from China tend to show more trust in robots and more acceptance of robots' recommendations while making decisions than people from South Korea and Germany (Li et al., 2010; Rau et al., 2009). The widely used Chinese human-automation trust scale (C-HATS) in China is the Chinese version of Jian et al., (2001)'s scale. Although reasonable psychometric properties of the Chinese version have been verified to be consistent with the original English version (Cai et al., 2022), as mentioned, Jian et al., (2001)'s scale cannot rest on any theory models. Another C-HATS was developed by Chien et al., (2014). The scale was developed based on theories, but as mentioned the contents of the items should be reconsidered for further applications in human-automation trust assessment. Consequently, there is still lack of an effective theoretically driven C-HATS.

### *1.3. Purpose of this study*

As outlined, Lee and See (2004)'s human-automation trust model better conceptualized human-automation trust and provided a more comprehensive understanding of human-automation trust. However, due to the dynamic and cross-cultural properties of trust, an effective measurement tool is necessary to further examine Lee and See (2004)'s model in various consumer automation product at different trust development stages in different cultural contexts. With this background, we aim to develop a C-HATS based on Lee and See (2004)'s model and test the reliability and validity of it for trust in different consumer automation products during different development stages. To better achieve our aims, there were three phases in the current study (**Table 1**). The data collection in the three phases were all approved by the ethical review board of the Faculty of Engineering, Information and Systems at the University of Tsukuba (approval number 2022R646)

|  | **Goals** | **Target system** | **Target trust** | **Specific systems** | **Methods** |
| --- | --- | --- | --- | --- | --- |
| **Phase 1** | Generate initial items | Driving related system | Initial trust | HADS | Content and face validity assessments; EFA |
| **Phase 2** | Verify whether the scale can be used for different trust stage. | Driving related system | Initial trust | CADS | EFA |
|  |  |  | Post-task trust | MOB GPS | EFA |
| **Phase 3** | Verify whether the scale can extend to other systems. | Non-driving related system | Initial trust | Robovac FRPS | IA; CFA; Reliability assessments |
|  |  |  | Post-task trust | FA stereo-garage ATC clothing |  |

**Table 1** Three phases of the development and assessment of the C-HATS. HADS: highly automated driving system; CADS: conditionally automated driving system; MOB GPS: mobile phone GPS navigation system; RV: robot vacuum; FRPS: facial recognition payment system; FA stereo-garage: fully automated stereo-garage; ATC clothing: automated temperature control clothing; EFA: exploratory factor analysis; IA: item analysis; CFA: confirmatory factor analysis. All adopted methods were referenced from Aiken and Groth-Marnat (2005).

Combing Phases 1 and 2, the dynamic differences of trust assessments at different trust development stages were verified. However, we examine the scale for initial and post-task trust only but not for dispositional and ongoing trust. The reason is that dispositional trust forms before recognizing a specific automation system and is less directly related to the concrete automation system, whereas ongoing trust is difficult to test by questionnaires during the interaction. As for the specific systems, HADS and CADS (SAE International, 2021) that have not come into the market yet, and MOB GPS that has already been experienced by most of the public in China were respectively used for initial and post-task trust. Besides, since the differences between the assessments on initial and post-task trust were few mentioned in previous studies (Razin, & Feigh, 2023), we presumed that the assessment items could be the same for initial and post-task trust and thus randomly chose initial trust to generate the initial assessment items in Phase 1. Then to examine the applicability of Phase 1's results on different automation system, a different driving automation system from the one in Phase 1 were used for Phase 2.

Then with Phase 3, the extendability of the developed C-HATS across various automation systems was further verified. Since systems chosen in Phase 1 and 2 were all driving related systems, non-driving related system such as FA stereo-garage and ATC clothing that has not come into the market yet, robovac and FRPS that have already been experienced by most of the public in China were chosen as the survey objects. In each phase, the actual used rates of automation systems of our participants were calculated.

## 2. Phase 1: Initial version of C-HATS

### 2.1. Methods
#### 2.1.1. Initial Item Generation
The initial item pool consisting of 21 items was built based on Lee & See (2004)'s descriptions of human-automation trust and existing items in Muir & Moray (1996), Korber (2019), and Madsen & Gregor (2000)'s trust scales. Seven items addressed *performance*, seven items assessed *process*, and another seven items reflected *purpose*. Specifically, among items of performance, two items were set up to assess the predictability of automation's actions, three items were set up to test the ability of automation, and two items were set up to reflect the reliability of automation's actions. Then for *process*, two items directly reflected its definition, and five items assessed the dependability, openness, and understandability of automation that were deemed to be related to process (Lee & See, 2004). For *purpose*, we developed two items to reflect the basic definition, and five items for faith and benevolence as mentioned in 1.1 were also formed. Finally, according to the results of content and face validity assessments, the initial version of C-HATS was formed. The C-HATS was scored on a Likert scale of 7 points, from 1 = "Strongly disagree" to 7 = "Strongly agree" in all assessment processes in the current study.

##### 2.1.1.1. Content validity
Nine assessors from China assessed the content validity of the initial item pool. The nine assessors were one researcher and two Ph.D. students in human factors, one associate professor and three Ph.D. students in psychology, and two user research engineers. They were asked to finish a questionnaire with a 5-point Likert scale (1= strongly inappropriate, 5= strongly appropriate) to judge whether the items were appropriate for testing the corresponding dimensions based on their understanding of Lee & See (2004)'s model. Combining assessors' subjective comments, we made some revisions to the items, deleted an item related to *dependability*, and added an item reflecting *integrity* (the adherence of trustees' actions to a set of acceptable principles, Mayer et al. (1995)) that was also related to *process* (Lee & See 2004). According to the results of the second round of assessment, the updated 21 items were all reserved since their scores were over or around 4 (appropriate, $p = .132$).

##### 2.1.1.2. Face validity
Face validity was then assessed. Four Chinese master students in psychology filled out the updated 21 items and identified obscure words and sentences, which were then modified for the initial version of C-HATS. As in Appendix 1 and 2, the term of the system was used in the sentences but could be changed for specific usage.

#### 2.1.2. Data collection and analyses
411 questionnaires including the initial version of C-HATS were distributed through the questionnaire release platform PowerCX and 368 questionnaires were recovered (recovery rate = 89.54%). We randomly chose initial trust to first examine the dimensions and items of the proposed C-HATS. The HADS (used rate = 0.3%) was chosen as the survey object.

Before starting the scale, we briefly introduced HADS with the introduction in Taxonomy of Driving Automation for Vehicles (GB/T 40429-2021) in China (SAC, 2021). Then participants were asked whether they had experienced HADS so far and to briefly describe their specific experiences with HADS. In this way, we excluded 42 invalid participants that have almost developed ongoing trust in HADS (1 participant), obviously misunderstood the introduction of HADS (7 participants) or answered questions carelessly (34 participants). However, for participants who checked "have no experience of HADS", we cannot screen out the ones who misunderstood the introduction of HADS, but the shortest answer time (around 3 minutes based on the pilot answers by the authors) and answer pattern check were used to discriminate the careless answers.

Finally, all 326 participants (127 men and 199 women, aged between 18-70 years) who answered "have no experience of HADS" were reserved as valid participants (the effective rate = 79.32%), concluding 294 participants with Chinese driving licenses and 32 participants without Chinese driving licenses. To examine the dimensions and items of the proposed C-HATS, an EFA was then performed in SPSS 26.0.

### 2.2. Results

The result of EFA and reserved items are shown in **Table 2**. The value of Bartlett's Test of Sphericity was 4413.353 ($df = 210$, $p < .001$). The value of Kaiser-Meyer-Olkin Test of Sampling Adequacy (KMO) was 0.971. Thus, the data was suitable for EFA, for which Principal Component Analysis (PCA) was used to extract the factors with eigenvalues greater than 1. Since the hypothesized factors (*performance*, *process*, and *purpose*) were conceptually related (Lee & See, 2004), oblique rotation with Promax was performed. Then after two turns of analysis, different from our hypothesis and Lee and See (2004)'s model, only two factors were finally obtained and items 1, 5, and 10 that loaded on both factors 1 and 2 with the secondary loading greater than 0.40 (Costello & Osborne, 2005; Beavers et al., 2013) were deleted.

**Table 2** Results of EFA for initial trust in Phase 1 (loadings of less than .32 were suppressed for pattern matrix)

| Item | Pattern matrix Factor 1 | Pattern matrix Factor 2 | Structure matrix Factor 1 | Structure matrix Factor 2 | Communality |
|---|---|---|---|---|---|
| 2.Predictability2 | 0.656 | | 0.758 | 0.616 | 0.584 |
| 3.Ability1 | 0.754 | | 0.796 | 0.603 | 0.635 |
| 4.Ability2 | 0.659 | | 0.745 | 0.596 | 0.562 |
| 6.Reliability1 | 0.668 | | 0.781 | 0.640 | 0.622 |
| 7.Reliability2 | 0.656 | | 0.794 | 0.665 | 0.647 |
| 8.Definition_process1 | 0.766 | | 0.791 | 0.589 | 0.625 |
| 9.Definition_process2 | 0.662 | | 0.774 | 0.634 | 0.611 |
| 11.Openness | 0.808 | | 0.727 | 0.472 | 0.534 |
| 12.Integrity | 0.741 | | 0.769 | 0.575 | 0.592 |
| 13.Understandbility1 | 0.802 | | 0.765 | 0.529 | 0.587 |
| 14.Understandbility2 | 0.815 | | 0.756 | 0.508 | 0.575 |
| 15.Definition_purpose1 | 0.646 | | 0.725 | 0.577 | 0.532 |
| 16.Definition_purpose2 | 0.838 | | 0.776 | 0.521 | 0.606 |
| 17.Faith1 | | 0.810 | 0.539 | 0.776 | 0.603 |
| 18.Faith2 | | 0.793 | 0.587 | 0.802 | 0.643 |
| 19.Faith3 | | 0.533 | 0.71 | 0.768 | 0.639 |
| 20.Benevolence1 | | 0.898 | 0.547 | 0.823 | 0.683 |
| 21.Benevolence2 | | 0.688 | 0.571 | 0.741 | 0.584 |
| Eigenvalues | 9.81 | 1.03 | / | / | / |
| Explained variance | 0.54 | 0.06 | / | / | / |

### *2.3. Discussion*

The results showed that the structure of initial trust was different from Lee and See (2004)'s theoretical hypotheses that human-automation trust can be predicted from three factors of *performance*, *process*, and *purpose*. Items that should respectively belong to performance or process were combined into only one dimension. In other words, *performance* and *process* were not two different dimensions anymore but as a whole. Therefore, there were finally only two dimensions for initial trust, *performance + process* and *purpose*. However, it is still not sure whether this result can be extended to initial trust in other driving related automation systems beyond HADS, and whether the result can also be extended to post-task trust in driving related automation systems. That is why we started the Phase 2 of the study.

## 3. Phase 2: Dimension of C-HATS for initial and post-task trust

In Phase 2, we examined the dimension of the initial version of C-HATS for both initial and post-task trust in other driving related automation systems by performing EFA again.

### *3.1. Methods*

340 questionnaires including the initial version of C-HATS were distributed through the questionnaire release platform Credamo. Each questionnaire contains two C-HATSs to respectively assess participants' initial and post-task trust in two different automation systems. As mentioned in 1.4., CADS (used rate = 19.67%) was selected for initial trust, whereas MOB GPS (used rate = 97.67%) was selected for post-task trust.

Before starting each scale, a video around two minutes selected from social media websites in China was used to introduce the automation system. For CADS, the videos presented the main functions with 3D animations that did not involve any specific products. For MOB GPS, the videos introduced the main functions of a specific product because appropriate 3D introduction animations that did not involve any specific products could not be found. We thus extra clarified that the products presented in the videos were only the examples, to avoid participants answering on those specific products.

After each video, participants were asked to make sure their usage experience of the automation systems and write down the names or the brands of the specific products if they have used. Finally, questionnaires with one or more answers of nonexistent products or with obvious response patterns for one or more scales were excluded, and 300 participants (109 men and 191 women, aged from 18-60 years) were reserved as valid participants (the effective rate = 88.23%), concluding 286 persons with and 14 persons without Chinese driving licenses.

### 3.2. Results

EFA were respectively performed for initial and post-task trust in SPSS 26.0. The followings are the results.

#### 3.2.1. Dimensions of Initial trust based on EFA

For initial trust, the value of Bartlett's Test of Sphericity was 3189.131 ($df = 210$, $p < .001$), and the value of KMO was 0.963, and the data was thus suitable for EFA. Same with phase 1, PCA was used to extract the factors and Promax was performed. After deleting items 5, 11, 12, and 14 with the first factor loading less than .05 (Costello & Osborne, 2005; Beavers et al., 2013), two factors were finally obtained, which was consistent to the result of phase 1. Then since item 2 did not load on the factor as hypothesized, it was also deleted. The final EFA results are shown in **Table 3**.

**Table 3** Results of EFA for initial trust in Phase 2 (loadings less than .40 were suppressed for pattern matrix)

| Item | Pattern matrix Factor | | Structure matrix Factor | | Communality |
|---|---|---|---|---|---|
| | 1 | 2 | 1 | 2 | |
| 1.Predictability1 | 0.643 | | 0.736 | 0.570 | 0.553 |
| 3.Ability1 | 0.627 | | 0.722 | 0.562 | 0.532 |
| 4.Ability2 | 0.736 | | 0.747 | 0.510 | 0.558 |
| 6.Reliability1 | 0.538 | | 0.735 | 0.655 | 0.588 |
| 7.Reliability2 | 0.532 | | 0.724 | 0.643 | 0.569 |
| 8.Definition_process1 | 0.801 | | 0.682 | | 0.483 |
| 9.Definition_process2 | 0.742 | | 0.720 | | 0.519 |
| 10.Dependability | 0.684 | | 0.730 | 0.528 | 0.536 |
| 13.Understandbility1 | 0.611 | | 0.685 | 0.520 | 0.476 |
| 15.Definition_purpose1 | 0.515 | | 0.641 | 0.533 | 0.430 |
| 16.Definition_purpose2 | 0.849 | | 0.713 | | 0.531 |
| 17.Faith1 | | 0.738 | 0.555 | 0.778 | 0.607 |
| 18.Faith2 | | 0.969 | | 0.829 | 0.712 |
| 19.Faith3 | | 0.516 | 0.612 | 0.694 | 0.521 |
| 20.Benevolence1 | | 0.88 | 0.541 | 0.846 | 0.717 |
| 21.Benevolence2 | | 0.754 | 0.549 | 0.783 | 0.614 |
| Eigenvalues | 7.72 | 1.23 | / | / | / |
| Explained variance | 0.48 | 0.07 | / | / | / |

#### 3.2.2. Dimensions of Post-task trust based on EFA

For post-task trust, the data was also suitable for EFA. The value of Bartlett's Test of Sphericity was 1413.499 ($df = 210$, $p < .001$) and the value of KMO was 0.888. As before, PCA and Promax was performed again. However, different from initial trust, three factors were finally obtained after deleting items 1, 2, 3, 4, 8, 11, 12, 13, 15 and 19 with a first factor loading less than .05 (Costello & Osborne, 2005; Beavers et al., 2013). The final results of EFA for post-task trust are shown in **Table 4**.

**Table 4** Results of EFA for post-task trust in Phase 2 (loadings less than .40 were suppressed for pattern matrix)

| Item | Pattern matrix Factor | | | Structure matrix Factor | | | Communality |
|---|---|---|---|---|---|---|---|
| | 1 | 2 | 3 | 1 | 2 | 3 | |
| 5.Ability3 | | 0.655 | | 0.480 | 0.741 | | 0.577 |
| 6.Reliability1 | | 0.616 | | | 0.657 | | 0.450 |
| 7.Reliability2 | | 0.882 | | | 0.829 | | 0.699 |
| 9.Definition_process2 | | | 0.663 | | | 0.700 | 0.516 |
| 10.Dependability | | | 0.591 | | | 0.622 | 0.464 |
| 14.Understandbility2 | 0.609 | | | 0.535 | | | 0.382 |
| 16.Definition_purpose2 | | | 0.755 | | | 0.735 | 0.553 |
| 17.Faith1 | 0.757 | | | 0.735 | | | 0.543 |
| 18.Faith2 | 0.617 | | | 0.697 | 0.498 | | 0.557 |
| 20.Benevolence1 | 0.627 | | | 0.686 | | | 0.488 |
| 21.Benevolence2 | 0.757 | | | 0.735 | | | 0.484 |
| Eigenvalues | 3.35 | 1.30 | 1.06 | / | / | / | / |
| Explained variance | 0.30 | 0.12 | 0.10 | / | / | / | / |

*3.3. Discussion*

In this phase, we further revised the initial version of C-HATS and respectively formed the final version of C-HATS for initial and post-task trust after combining the results from phases 1 and 2. For initial trust, considering the simplicity of the scale, items 1, 2, 5, 10, 11, 12, and 14 that were deleted in phases 1 and/or 2 were all deleted, and the final version of C-HATS concludes 14 items. For post-task trust, as in phase 2, there were 11 items reserved in the final version of C-HATS.

Consistent with the results of Phase 1, the results of Phase 2 on different automation systems further showed that the structure of initial trust should be different from Lee and See (2004)'s theoretical hypotheses, consisting of only two dimensions, *performance + process* and *purpose*. Whereas, the results of Phase 2 also showed the different structures of initial and post-task trust. For post-task trust, there should be three dimensions just as Lee and See (2004)'s model proposed.

Nevertheless, both Phase 1 and 2 were performed on driving-related system only. As mentioned in introduction, the suitability of C-HATS for various consumer automation products is necessary. Consequently, the verification of our C-HATS on driving-related automation system in Phase 1 and 2 are still not enough, and Phase 3 were performed to further verify our C-HATS on different non-driving related automation system.

# 4. Phase 3: Reliability and validity of C-HATS

In this phase, we aim to check the extendability of the final version of C-HATS across other non-driving related automation systems. The followings are the methods and results of the reliability and validity assessments respectively for initial and post-task trust in other automation systems.

### 4.1. Methods

The 340 participants in phase 2 also finished the other four C-HATSs, of which two were for initial trust in FA stereo-garage (used rate = 8.67%) and ATC clothing (used rate = 13.67%) and another two were for post-task trust in robovac (used rate = 79.00%) and FRPS (used rate = 84.33%). These responses (response 1) were used in this section to assess the item discrimination, reliability and validity of the final version of C-HATS.

Before answering the questionnaires, participants were introduced of the systems with videos around two minutes selected from social media websites. For FA stereo-garage, the main functions were presented with 3D animations that did not involve any specific products. For ATC clothing, robovac, and FRPS, the main functions were introduced via four specific products, for which clarifications were also made to avoid participants answering on the specific products. After each video, the confirmations of participants' usage experiences of the systems were also done. Finally, there were 300 valid participants (109 men and 191 women, aged from 18-60 years, the effective rate = 88.23%) concluding 286 persons with and 14 persons without Chinese driving licenses, of which 243 persons having used none of the two systems for initial trust, and 199 persons having used both the two systems for post-task trust.

#### 4.1.1. Item discrimination assessment

To assess item discrimination, IA of the final version of C-HATS were respectively performed for initial and post-task trust in SPSS 26.0. First, the average scores of the two systems on all the items were calculated to be used for the later analyses. According to the sum score of the whole scale (14 items for initial trust and 11 items for post-task trust), the valid participants were sorted and the top 27% and the bottom 27% were respectively taken as high-trust and low-trust group. Then independent-sample t-tests on all item scores were performed between the high and low-trust groups to determine whether the difference of the mean scores (the Critical Ration, CR) between the high and low-trust groups in each item was significant (Yuan et al., 2019). Finally, the Item-Total Correlation (ITC) values between all item scores and the sum score of the scale were also calculated, for which items with a low coefficient ($r < 0.4$) and/or a high $p$ value ($p > 0.05$) should be removed (Lee & Nicewander, 1988).

#### 4.1.2. Validity assessment

To assess the construct validity of the final version C-HATS, CFA was respectively performed for initial and post-task trust in Amos 26.0. A good construct validity needs a good model fit in CFA, which is commonly indicated by Chi-square/df, the goodness of fit index (GFI), adjusted goodness of fit index (AGFI), comparative fit index (CFI), and root mean square error of approximation (RMSEA). According to Hair et al. (2010), Chi-square/df should be less than 5 (3 if strictly), GFI, AGFI, and CFI should be greater than 0.9, and RMSEA should be less than 0.5.

### 4.1.3. Reliability assessment

The internal consistency reliability, split-half reliability and test-retest reliability assessments were then performed in SPSS 26.0. For internal consistency reliability, Cronbach's coefficient α was calculated, which should be greater than 0.7. For split-half reliability, the Guttman split-half coefficient was calculated and should also be greater than 0.7 (Cronbach, 1951). To assess the test-retest reliability, an additional retest of the final version of C-HATS were performed 32 days after collecting response 1. 149 participants (44 men and 97 women, aged 18-60 years) from the valid 300 participants of response 1 were randomly selected. Then, using the same method as mentioned in response 1, careless participants were removed, and there were finally 141 valid participants (the effective rate = 94.63%), among which 100 participants had used none of the two systems and were selected for initial trust, whereas 119 participants had used both the two systems and were selected for post-task trust. Then the scores by the retest of the total scale and each dimension were matched to the scores of response 1, and correlation analyses were performed between the matched scores, for which a greater correlation coefficient means a higher test-retest reliability (Wu et al., 2019) and 94 participants for initial trust and 101 participants for post-task trust were finally matched successfully.

### 4.2. Results
### 4.2.1. Item discrimination

The results of independent-sample t-tests showed that the CR values were significant on all items for both initial and post-task trust between high and low-trust groups ($ps < .001$). In addition, significant correlations were found on all items for both initial ($rs \geq .52$, $ps < .01$) and post-task trust ($rs \geq .40$, $ps < .01$). Consequently, it is considered that the final version of C-HATS has good item discrimination for both initial and post-task trust. The CR values and the ITC values are shown in **Table 5**.

**Table 5** CR values and item-total correlation values for item analysis

| Item | Initial trust | | Post-task trust | |
|---|---|---|---|---|
| | *CR* | *r* | *CR* | *r* |
| 1.Predictability1 | / | / | / | / |
| 2.Predictability2 | / | / | / | / |
| 3.Ability1 | -13.21 | 0.70** | / | / |
| 4.Ability2 | -12.38 | 0.67** | / | / |
| 5.Ability3 | / | / | -9.67 | 0.73** |
| 6.Reliability1 | -11.18 | 0.67** | -9.28 | 0.69** |
| 7.Reliability2 | -14.10 | 0.70** | -11.46 | 0.76** |
| 8.Definition_process1 | -11.97 | 0.64** | / | / |
| 9.Definition_process2 | -9.77 | 0.60** | -7.07 | 0.51** |
| 10.Dependability | / | / | -7.66 | 0.49** |
| 11.Openness | / | / | / | / |
| 12.Integrity | / | / | / | / |
| 13.Understandbility1 | -11.81 | 0.68** | / | / |
| 14.Understandbility2 | / | / | -9.39 | 0.62** |
| 15.Definition_purpose1 | -10.40 | 0.63** | / | / |
| 16.Definition_purpose2 | -7.47 | 0.52** | -6.92 | 0.40** |
| 17.Faith1 | -12.92 | 0.68** | -10.62 | 0.76** |
| 18.Faith2 | -12.96 | 0.69** | -9.00 | 0.76** |
| 19.Faith3 | -12.00 | 0.67** | / | / |
| 20.Benevolence1 | -10.63 | 0.60** | -10.36 | 0.77** |
| 21.Benevolence2 | -10.97 | 0.63** | -9.36 | 0.70** |

**The correlation was significant at 0.01 level. "/" means the value was not calculated in IA, since the item was deleted based on the results from phases 1 and 2.

### 4.2.2. Validity

The model fit indices showed reasonable fits to data of both initial trust ($\chi^2_{(76)}$ = 127.54, GFI = .931, AGFI = .904, CFI = .958, RMSEA = .053) and post-task trust ($\chi^2_{(41)}$ = 87.69, GFI = .918, AGFI = .868, CFI = .943, RMSEA = .076). The standardized factor loading of the items was all greater than 0.5 ($ps$ < .01) (**Figures 2 and 3**). These indicate that the C-HATS has acceptable construct validity.

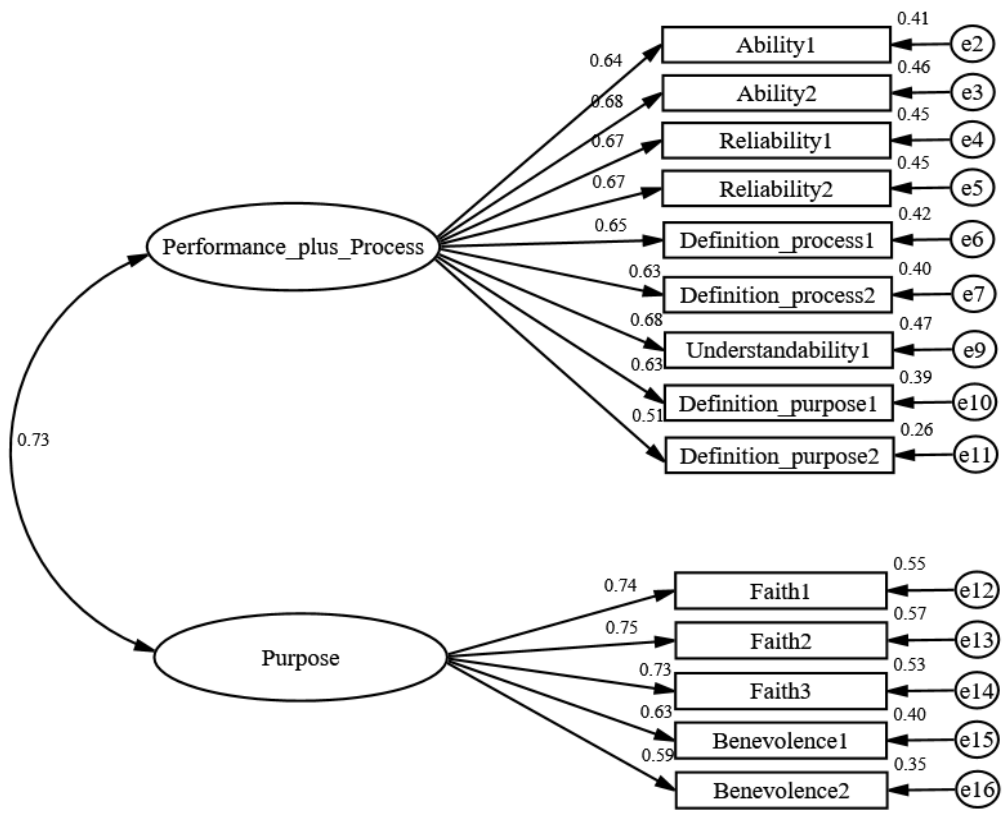

**Figure 2** Results of CFA for initial trust.

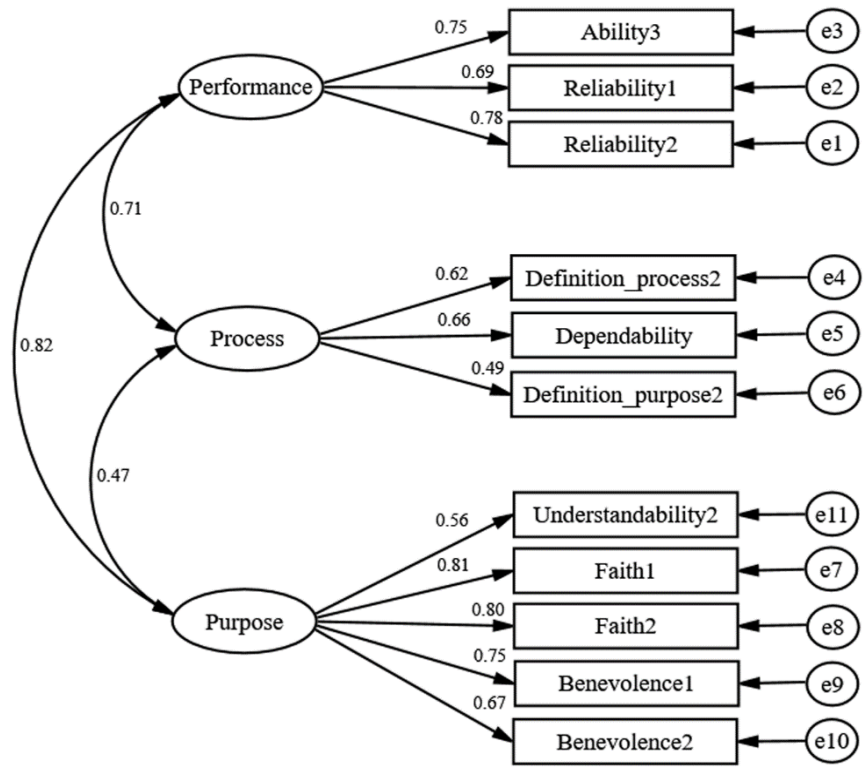

**Figure 3** Results of CFA for post-task trust.

### *4.2.3. Reliability*

The reliability indexes are shown in **Tables 6 and 7**.

For initial trust, the results showed that Cronbach's coefficient alpha values and the split-half coefficients of the total scale and each dimension of the scale were all greater than 0.7 (see **Table** 6). This indicates that the total scale and each dimension of the final version of C-HATS for the initial trust had good internal consistency. In addition, the correlation coefficients of the matched scores were greater than 0.65 ($ps < 0.01$), indicating acceptable stability across time of the scale.

For post-task trust, Cronbach's coefficient alpha values and the split-half coefficients of the total scale and the dimensions of the scale were greater than 0.7 (see **Table** 7), besides the dimension of process. Therefore, it is considered that, besides the dimension of process, the total scale and the dimension of performance and purpose had good internal consistency. Then, besides the dimension of process ($r = 0.55$), the correlation coefficients of the matched scores were all greater than 0.7 ($ps < 0.01$), indicating acceptable stability across time of the total scale and the dimension of performance and process.

**Table 6** Results of reliability assessments for initial trust.

|  | Cronbach's coefficient α | Split-half coefficient | Test-retest coefficient |
|---|---|---|---|
| Performance + Process | 0.86 | 0.85 | 0.67** |
| Purpose | 0.82 | 0.73 | 0.72** |
| Total | 0.89 | 0.84 | 0.75** |

***The correlation was significant at 0.001 level.

Table 7 Results of reliability assessments for post-task trust.

|  | Cronbach's coefficient α | Split-half coefficient | Test-retest coefficient |
|---|---|---|---|
| Performance | 0.78 | 0.71 | 0.70** |
| Process | 0.62 | 0.50 | 0.55** |
| Purpose | 0.84 | 0.73 | 0.81** |
| Total | 0.87 | 0.69 | 0.82** |

***The correlation was significant at 0.001 level.

### 4.3. Discussion

The results in this phase on different non-driving related automation systems further showed the consistencies of the results in Phase 1 and 2 across various automation systems. Therefore, for various consumer automation products, post-task trust should have three dimensions as Lee and See (2004)'s model, whereas initial trust should have only two dimensions, *performance + process* and *purpose*. In addition, the results in this phase verified that the final version of C-HATS had reasonable reliability and validity for human-automation trust in different consumer automation products during different development stages. The following section will summarize and further discuss the above findings in the three phases and integrate them with those of previous studies to shed light on the practices of human-automation trust assessment.

# 5. Discussions

### 5.1. Development of C-HATS

In the current study, we developed and tested a reliable and valid human-automation trust scale in Chinese cultural context. For both initial and post-task trust assessments, all the items had good discrimination since the scores of the high-score group were all significantly greater than those of the low-score group and the total correlations of items were equal to or greater than 0.59. However, it should be noted that the final version of C-HATSs for initial and post-task trust assessments have different dimensions and items based on the results of EFA. For the post-task trust, the scale has three dimensions and 11 items, of which all factor loading values were over 0.50 and the explanation rate reached 51.93%. This reflected the consistency of our results with Lee and See (2004)'s model. Whereas, for initial trust, the final scale had 14 items with factor loading values over 0.50 but only two dimensions with an explanation rate of 55.90%, which was different from Lee and See (2004)'s model and will be further discussed in the next sections. Nevertheless, no matter initial or post-task trust, the results of CFA indicated the reasonable construct validity of C-HATS. In addition, by combining the values of Cronbach's coefficient α, split-half coefficient, and test-retest coefficient, both the initial and post-task trust scale have good reliability. In Appendix 1, we provide all 21 developed items and show which were finally reserved for initial and post-task trust.

### 5.2. Application of C-HATS

Besides assessments of the psychometric properties of the final version of C-HATS, we also verified the applicability of C-HATS on assessing human-automation trust in different consumer automation products during different development stages.

*5.2.1. Application of C-HATS at different trust development stages*

Combining the results of the three phases, it was known that the structures of human-automation trust before and after the interactions were not totally the same. Same with Lee and See (2004)'s model, post-task trust can be predicted by three factors of *performance*, *process*, and *purpose*, whereas for initial trust there were only two dimensions, *performance + process* and *purpose*. Therefore, human-automation trust at different development stages should be respectively considered and tested with different items. Some items of our C-HATS are not appropriate for both initial and post-task trust assessments, and should thus be respectively used for initial and post-task trust.

Based on the current literature, human-automation trust can also be divided into *cognition-based* and *affect-based trust* (Madsen & Gregor, 2000; Glikson & Woolley, 2020). *Cognition-based trust* is grounded in evidence of the "trustworthiness" or "controllability" (recently proposed by Kieseberg et al. (2023)) of a system according to trustors' prior cognition, whereas *affect-based trust* relies on trustors' attributions concerning the motives of trustees (Lewis & Weigert, 1985). In the current study, in the context of initial trust, considering the connotations of the definitions and items of *performance* and *process*, *performance* and *process* are both cognition-related evidence of "trustworthiness" or "controllability", and can thus be combined into one dimension named *cognition-based trust*. Similarly, *purpose* is similar to *affect-based trust* including *faith* and *benevolence*. However, it should also be noted that *purpose* purposed by Lee and See (2004) should be further divided into two specific parts, recognition of the actual design intents of a system (eg. finishing a certain task) and perception of the positive orientation of a system with emotion in an anthropomorphic way (eg. hurt me or not). The former should be one cognitive trust-building process of cognition-based trust, whereas the latter should be considered as affect-based trust. Based on our results of EFA, our items 15 and 16 reflecting Lee and See (2004)'s definition of *purpose* were not put together with other items (items 17-21) related to *faith* and *benevolence* that were considered to correspond to *purpose* (Lee & See, 2004). According to the connotations, our items 15 and 16 conveyed the participants' recognition of the system's actual design intents, whereas items 17-21 reflected participants' emotional perception of the positive orientation of the systems.

*5.2.2. Application of C-HATS in different consumer automation products*

The results showed that the final version of C-HATS can be used to assess initial and post-task trust in various consumer automation product systems. The structures of initial and post-task trust were consistent across driving related and non-driving related automation systems. Only one thing that should be noted is the difference between the structures of initial and post-task trust. Actually, another possible factor that enhance the difference in the structures of initial and post-task trust may be the consumer attributes of the selected automation systems in the current study. Prior cognition of an emerging automation system before interactions plays a crucial role in building appropriate initial trust in the system, since it can to some extent influence both the level (Kim et al., 2004) and the structure (Lee et al., 2021) of the initial trust in the system. For non-consumer automation product, such as cockpit automation, professional and even tailored pilot training curricula before the real pilot-automation interaction were con-sidered (Sarter & Woods, 1992). However, for consumer automation product, although the participants may have more or less learn something related to the systems, most of them would not been professionally educated and trained on how the systems work and how to use and interact with the systems. The participants thus have less or even few of concrete cognition of the systems.

We also found that when testing users' trust in a certain automation system, the items should also be properly added on or subtracted from our C-HATS. Based on the results, although the dimensions of initial trust were same for different systems, the specific deleted items were different in Phase 1 and 2. That is, not all specific items can reflect human trust in all kinds of systems. To better assess human-automation trust in a certain system, some items that are not suitable for the system should be eliminated, whereas some other new items that can better reflect user's trust in the system should be added.

### 5.3. *Theoretical Implications*

Combing previous theories on human trust (eg., Lee & See, 2004; Lewis & Weigert, 1985; Glikson, & Woolley, 2020; Holzinger, 2021) and the results of the current study, there were several theoretical implications for human-automation trust theories. Firstly, as mentioned in 5.2.1, same as interpersonal trust (eg., Lewis & Weigert, 1985) and human-computer trust (Madsen and Gregor, 2000), human-automation trust could also have two parts - cognition-based trust and affect-based trust. Cognition-based trust depends on objective and rational evidence whereas affect-based trust relies on emotional factors. Then, cognition-based trust of human-automation trust can be built through several subordinate cognitive processes. As Doney et al. (1998)'s study on trust development in business contexts, cognitive trust-building processes conclude calculating the benefits of trust, predicting trustees' behaviors, evaluating trustees' motivations, assessing trustees' ability, and transferring trust from other entities related to trustees to the trustees. However, according to different automation systems and different cognition levels of the users, cognitive processes that will be used should be different. The major cognitive processes can be related to the Lee and See (2004)'s *performance, process*, and recognition parts of *purpose* on actual design intents of a system, which also respectively include several facets. Since the facets are not completely independent, some facets will not clearly belong to the only one cognitive process. This may also be the reason why items 14 and 16 for post-task trust were not clearly put into the hypothesized dimension.

### 5.4. *Limitations*

This study has several limitations. First, due to the difficulty to find kinds of consumer automation product systems that possess almost equal number of users who have used it and have not used it, assessments of initial and post-task trust were not performed on the same systems. In addition, since hundreds participants were needed to do EFA and CFA, it was also difficult to collect inexperienced participants to experience certain automation systems and test their initial and post-task trust before and after some interactions. Second, we could not directly reveal the reasons why the structures of initial and post-task trust were different, for which future studies can continue to explore. Finally, future studies should be conducted with a larger sample size on more automation systems to further verify the reliability and validity of the developed C-HATS.

# 6. Conclusion

Based on Lee and See (2004)'s trust model and existing trust scales, a Chinese version of human-automation trust scale (C-HATS) was developed. For human-automation trust at initial and post-task stages, different dimensions and items were considered. For the post-task trust, the scale had three dimensions and 11 items and reflected Lee and See (2004)'s model, whereas different from Lee and See (2004)'s model, the final scale had 14 items but only two dimensions for initial trust. Nevertheless, for both initial and post-task trust, reasonable reliability and validity of the scale were verified with various consumer automation products, although further verification is still necessary.

# Acknowledgments

This work was supported under the framework of international joint research program managed by the National Research Foundation of Korea (NRF-2022K2A9A2A08000167), and the Japan Society for the Promotion of Science (JPJSBP 120228804). We thank Hang Yuan, Huiping Zhou, Jianxin Ou, Liangjie Chen, Mengjiao Liu, Nianzhi Tu, Shuying, Li, Suyang An, Xiaodi Wu, and Yuan Zhang's help in content validity assessments and item development.

# Appendix 1 Items of C-HATS

**xxx 自动化系统信任问卷（C-HATS）**

以下是关于 xxx 自动化系统的一些描述，请您针对 xxx 自动化系统，从 1(非常不同意)到 7(非常同意)中选择与您的真实想法相符的数字。

The following is a description of the xxx automation system. Please select the number from 1(Strongly disagree) to 7(Strongly agree) that corresponds to your true opinion of the xxx automation system.

| 1 非常不同意　2 不同意　3 比较不同意　4 一般　5 比较同意　6 同意　7 非常同意 | Initial dimension assignment (Lee & See, 2004) | Keywords (References) | Initial item pool | Initial version | Final version (Initial trust) | Final version (Post-task trust) |
|---|---|---|---|---|---|---|
| 1. xxx 系统的性能是稳定的。/ The performance of the system is stable. | Performance | Predictability (Lee & See, 2004) | √ | √ | | |
| 2. xxx 系统的行为表现是可以预测的。/ The behavior of the xxx system is predictable. | Performance | Predictability (Lee & See, 2004) | √ | √ | | |
| 3. xxx 系统总能为我提供合理的帮助或建议。/ The xxx system can always provide me with reasonable help or advice. | Performance | Ability (Lee & See, 2004) | √ | √ | √ | |
| 4. xxx 系统能够恰当地执行各种预设功能。/ The xxx system can properly perform various preset functions. | Performance | Ability (Lee & See, 2004) | √ | √ | √ | |
| 5. xxx 系统能够始终如一地应对预设的各种问题。/ The xxx system is able to consistently respond to preset problems. | Performance | Ability (Lee & See, 2004) | √ | √ | | √ |
| 6. xxx 系统的行为表现是可靠的。/ The behavior of the xxx system is reliable. | Performance | Reliability (Lee & See, 2004) | √ | √ | √ | √ |
| 7. xxx 系统总能为我提供真实可靠的信息。/ The xxx system can always provide me with true and reliable information. | Performance | Reliability (Lee & See, 2004) | √ | √ | √ | √ |
| 8. xxx 系统的设计适合预设的应用场景。/ The design of the xxx system is suitable for preset application scenarios. | Process | Definition of process (Lee & See, 2004) | √ | √ | √ | |

| 题项 / Item | 维度 | 来源 | | | | |
|---|---|---|---|---|---|---|
| 9. xxx 系统的设计符合预期的使用目的。/ The design of the xxx system conforms to my intended purpose of use. | Process | Definition of process (Lee & See, 2004) | √ | √ | √ | √ |
| 10.0 xxx 系统自始至终都是值得依赖的。/ The xxx system is dependable. | Process | Dependability (Muir & Moray, 1996) | √ | | | |
| 10. 我能够依赖 xxx 系统以达到我的使用目的。/ I can rely on the xxx system for my purposes of use. | Process | Dependability (Muir & Moray, 1996) | √ | √ | | √ |
| 11. xxx 系统会在操作过程中与我进行清晰明确的交流和沟通（如，向我反馈信息并听取我的意见或观点）。/ The xxx system will clearly communicate with me during the operation (e.g., give me feedback and listen to my opinions). | Process | Openness (Schindler & Thomas, 1993) | √ | √ | | |
| 12. 我相信 xxx 系统在操作过程中会一直遵循其设计原则（如，保证人机安全）。/ I believe that the xxx system will always adhere to a set of design principles (e.g., to ensure human and machine safety). | Process | Integrity (Mayer, 1995) | | √ | | |
| 13. 今后使用 xxx 系统时，我知道我应该做什么才能从该系统获得所需支持。/ When using the xxx system in the future, I know what I should do to get the supports I need from it. | Process | Understandability (Sheridan, 1992) | √ | √ | √ | |
| 14. 我目前能够理解 xxx 系统的设计原理和工作机制。/ I can now understand the design principles and mechanisms of the xxx system. | Process | Understandability (Sheridan, 1992) | √ | √ | | √ |
| 15. 我目前很清楚 xxx 系统的设计目标和功能范围。/ I now have a clear idea of the realm of the designer's intent of the xxx system. | Purpose | Definition of purpose (Lee & See, 2004) | √ | √ | √ | |
| 16. 我会在 xxx 系统的功能范围内和适用条件下使用该系统。/ I will use the xxx system within its functional limits and under its applicable conditions. | Purpose | Definition of purpose (Lee & See, 2004) | √ | √ | √ | √ |

| Item | | Keyword | | | | |
|---|---|---|---|---|---|---|
| 17.即使我不清楚解决某一个难题是否属于 xxx 系统的设计目标和功能范围，我也会尝试使用该系统去解决这个难题。/ Even if I am not sure if solving a problem is within the design intents and the capabilities of the xxx system, I will try to use the system to solve the problem. | Purpose | Faith (Muir & Moray, 1996) | √ | √ | √ | √ |
| 18. 当我不确定 xxx 系统在某个应用场景中是否可用时，我仍然选择相信该系统在这个应用场景中能为我提供合理的帮助或建议。/ When I am not sure if the xxx system is available in an application scenario, I still believe that the system can provide me with reasonable help or advice in that application scenario. | Purpose | Faith (Muir & Moray, 1996) | √ | √ | √ | √ |
| 19. 当我对某个决策没把握时，我相信 xxx 系统能为我提供最佳方案。/ When I'm not sure about a decision, I believe the xxx system to provide the best solution for me. | Purpose | Faith (Muir & Moray, 1996) | √ | √ | √ | |
| 20. 即使 xxx 系统存在一些系统漏洞，我仍相信该系统在操作过程中不会出现问题，更不会伤害到我。/ Even if there are some bugs in the system, I still believe that the xxx system will behave without problems and will not hurt me. | Purpose | Benevolence (Mayer, 1995) | √ | √ | √ | √ |
| 21. xxx 系统出现一些错误也是情有可原的，因为我相信该系统设计的初衷总归是对我好的。/ It's pardonable that some errors occur in the system because I believe the xxx system was designed to be good for me. | Purpose | Benevolence (Mayer, 1995) | √ | √ | √ | √ |

The references in the column of keywords show the references that described the keywords, based on which we developed the items. "√" means that the item exited in the initial item pool, the initial version of the C-HATS, or the final version of the C-HATS on initial trust or post-task trust.